\documentclass[11pt,a4paper]{article}

\usepackage[margin=1in]{geometry}
\usepackage{amsmath,amssymb}
\usepackage{graphicx}
\usepackage{booktabs}
\usepackage{multirow}
\usepackage{array}
\usepackage[colorlinks=true,linkcolor=blue,citecolor=blue,urlcolor=blue]{hyperref}
\usepackage[numbers,sort&compress]{natbib}
\usepackage{xcolor}
\usepackage{caption}
\usepackage{subcaption}
\usepackage{tikz}
\usetikzlibrary{arrows.meta,positioning,shapes.geometric,fit,backgrounds}
\usepackage{enumitem}
\usepackage{parskip}
\usepackage{authblk}
\usepackage[font=small]{caption}

\makeatletter
\def\fps@figure{ht}
\def\fps@table{ht}
\makeatother

\newcommand{\msai}{MarketSenseAI}
\newcommand{\vt}{\mathbf{t}}
\newcommand{\va}{\mathbf{a}}
\newcommand{\vw}{\mathbf{w}}
\newcommand{\wh}{\hat{w}}
\newcommand{\IC}{\ensuremath{\mathrm{IC}}}
\newcommand{\ICIR}{\ensuremath{\mathrm{ICIR}}}
\newcommand{\RR}{\mathbb{R}}
\newcommand{\spx}{S\&P~500}
\newcommand{\spone}{S\&P~100}

\author{George Fatouros}
\author{Kostas Metaxas}
\affil{Alpha Tensor Technologies\\
  \href{mailto:george@alpha-tensor.ai}{\texttt{george@alpha-tensor.ai}}
  \quad
  \href{mailto:kostas@alpha-tensor.ai}{\texttt{kostas@alpha-tensor.ai}}}

\title{\textbf{Signal or Noise in Multi-Agent LLM-based\\
Stock Recommendations?}\\[0.5em]
}

\date{April 2026}

\begin{document}
\maketitle

\begin{abstract}
We present the first portfolio-level validation of MarketSenseAI\footnote{\url{https://marketsense-ai.com}}, a deployed multi-agent LLM equity system.  All signals are generated live at each observation date, eliminating look-ahead bias.  The system routes four specialist agents---News, Fundamentals, Dynamics, and Macro---through a synthesis agent that issues a monthly equity thesis and ordinal recommendation for each stock in its coverage universe, and we ask two questions: do its buy recommendations add value over both passive benchmarks and random selection, and what does the internal agent structure reveal about the source of the edge?

On the \spx{} cohort (19 months) the strong-buy equal-weight portfolio earns $+2.18\%$/ month against a passive equal-weight benchmark of $+1.15\%$ (approximating RSP\footnote{\href{https://www.invesco.com/us/en/financial-products/etfs/invesco-sp-500-equal-weight-etf.html}{Invesco S\&P 500 Equal Weight ETF}}), a $+25.2$ percentage-point compound excess, and ranks at the 99.7th percentile of 10{,}000 Monte Carlo null portfolios ($p=0.003$).  The \spone{} cohort (35 months) delivers a $+30.5$ percentage-point compound excess over EQWL\footnote{\href{https://www.invesco.com/us/en/financial-products/etfs/invesco-sp-100-equal-weight-etf.html}{Invesco S\&P 100 Equal Weight ETF}} with consistent direction but formal significance not reached ($p=0.17$), limited by the small average selection of ${\sim}10$ stocks per month.

Non-negative least-squares (NNLS) projection of thesis embeddings onto agent embeddings reveals an adaptive-integration mechanism rather than a dominant-agent effect.  Agent contributions rotate with market regime
(Fundamentals leads on \spx{}, Macro on \spone{}, Dynamics acts as an episodic momentum signal) and this agent rotation moves in lockstep with both the sector composition of strong-buy selections and identifiable macro-calendar events---three independent views of the same underlying adaptation.  The ordinal recommendation's
cross-sectional Information Coefficient (\IC{}) is statistically significant on \spx{} ($\ICIR{}=+0.489$, $p=0.024$).  These results suggest that multi-agent LLM equity systems can identify sources of alpha beyond what classical factor models capture, and that the strong-buy signal functions as an effective universe-filter that can sit upstream of any portfolio-construction process.
\end{abstract}


\section{Introduction}

\subsection{Research Question and Motivation}

Multi-agent LLM systems are increasingly deployed in equity research, yet
rigorous out-of-sample evidence on whether their signals translate into
portfolio-level performance remains scarce.  This paper addresses one
focused question: \emph{do the strong-buy recommendations produced by a
deployed multi-agent LLM equity system beat random stock selection?}

We study \msai{}, a system that
synthesises specialist agent analyses---News, Fundamentals, Dynamics, and
Macro---into a unified thesis and ordinal recommendation for each stock.
We test performance using a Monte Carlo benchmark that matches the actual
portfolio on every dimension except selection: same universe, same dates,
same number of holdings, same equal weighting.  We then decompose thesis embeddings via Non-Negative Least Squares (NNLS) attribution to characterise
how the four agents contribute collectively, finding that their
contributions are heterogeneous and context-dependent---consistent with
the synthesis agent adaptively weighting agents by sector and market regime.

The paper makes no claim that LLM systems generically outperform markets.
It provides a transparent empirical case study of one deployed system,
including transparent reporting of where statistical evidence is and is not
sufficient to draw conclusions.

\subsection{Related Work}

\paragraph{LLMs in financial analysis:}
\citet{lopezlira2023chatgpt, fatouros2025can} show that general-purpose LLMs encode market-relevant information, while domain-adapted
models (BloombergGPT \citep{wu2023bloomberggpt}, FinGPT\citep{yang2023fingpt}) improve on financial tasks. \citet{fatouros2023transforming} and \citet{papasotiriou2024ai} establish that LLM-generated sentiment correlates with short-horizon equity returns.

\paragraph{Multi-agent and agentic systems:}
Recent work moves beyond single-model approaches. TradingAgents \citep{xiao2024tradingagents} and AlphaAgents \citep{zhao2025alphaagents} coordinate specialist LLM agents for portfolio construction. Mixture-of-experts architectures \citep{vallarino2025moe} dynamically route inputs across specialised sub-models, while regime-aware approaches \citep{zhang2025regimefolio} adapt factor weights to market conditions. Expert-team frameworks \citep{miyazaki2026expertteams} decompose complex financial tasks across LLM agents with structured synthesis.

\paragraph{LLM signal evaluation:}
A growing literature cautions against over-claiming.  \citet{li2025profitmirage}
identify data-snooping biases in LLM backtests; \citet{shergadwala2026stability}
document instability across prompts; \citet{almarri2025measuring} and
\citet{han2024aiagent} measure LLM overconfidence in financial contexts.
\citet{satapathy2025earnings} and \citet{geng2026could} assess LLM performance
on earnings forecasting.  \citet{fatouros2025marketsenseai} describe the
\msai{} system architecture evaluated here.

\paragraph{Embedding-based attribution:}
SPLICE \citep{bhalla2024splice} and Atlas \citep{papadakis2025atlas} apply
non-negative decompositions to financial text embeddings.
\citet{tatsat2025blackbox} provide interpretability methods for LLM-generated
financial signals.  Prompt sensitivity in financial embeddings is studied by
\citet{wang2024prompt} and \citet{kulpa2025promptfinance}.

\paragraph{Factor models and IC:}
Our attribution methodology connects to classical factor-model practice
\citep{grinold2000active,barra2011use4,fama1993common}, with the
information coefficient as a performance measure following
\citet{grinold2000active} and cross-sectional dependence corrections
following \citet{northfield2003crosssectional}.

\subsection{Contributions}

This paper makes three empirical contributions:

\begin{enumerate}
    \item \textbf{Portfolio selection skill (primary):}We provide the first Monte Carlo validation of a deployed multi-agent LLM equity system at the portfolio level, testing whether strong-buy selections outperform random same-sized portfolios drawn from the same universe on the same dates.  The Monte Carlo design controls for universe composition, date-timing, and position count; only stock selection differs.
    \item \textbf{Agent-level attribution (secondary):}NNLS decomposition of thesis embeddings recovers a panel of per-stock-date agent contribution weights, validated by cosine diagnostics.  Agent
contributions are heterogeneous and context-dependent: no single agent dominates across all dates and sectors, and the dominant agent rotates with market regime, consistent with the synthesis agent exploiting each agent's comparative advantage.
    \item \textbf{Implicit vs explicit recommendation signal (secondary):}We compare the predictive content of the continuous NNLS agent weights against the discrete ordinal recommendation within the actionable buy+strong-buy universe.  Both the explicit recommendation label and the implicit agent weights carry genuine predictive content, with the continuous weights encoding cross-sectional return information at a finer grain than the discrete label.
\end{enumerate}

The remainder of this paper is organised as follows.
Section~\ref{sec:system} describes the \msai{} system and the two fixed
cohorts.  Section~\ref{sec:methodology} develops the Monte Carlo portfolio
test, the NNLS attribution methodology, and the information-coefficient
framework.  Section~\ref{sec:data} presents descriptive statistics for the
signal panel.  Section~\ref{sec:results} presents the empirical results:
Monte Carlo selection skill, NNLS attribution diagnostics, risk and return
profile, and agent IC analysis, followed by sector heterogeneity.
Section~\ref{sec:discussion} discusses limitations and a market-beta
robustness check, and Section~\ref{sec:conclusion} concludes.

\section{System Overview}
\label{sec:system}

\msai{} is an operational
equity-research platform developed by Alpha Tensor
Technologies.\footnote{\url{https://www.alpha-tensor.ai}}
It ingests financial news, fundamentals, price dynamics, and macro data to produce a structured investment thesis and an ordinal recommendation for each stock in its coverage universe on a fixed frequency.

\paragraph{Four specialised agents:}
The four agents are: \textbf{News}~-- ticker-specific progressive news
analysis capturing recent company-level developments; \textbf{Fundamentals}~--
quantitative financials, filings and earning transcripts; \textbf{Dynamics}~-- price-action and technical signals; and
\textbf{Macro}~-- sector-level and macroeconomic context.

Each agent independently produces a text analysis for a given stock and date. A synthesis agent reads all four expert analyses and generates a free-text equity thesis together with a five-point ordinal recommendation:
\emph{strong sell}, \emph{sell}, \emph{hold}, \emph{buy}, \emph{strong buy}.
The ordinal recommendation is the synthesis agent's explicit sentiment
assessment of the stock: it distils the thesis into a discrete conviction
label ranging from the most bearish (\emph{strong sell}) to the most
bullish (\emph{strong buy}).  For quantitative analysis we encode the
recommendation as a numerical \emph{ordinal score}; within the actionable
buy+strong-buy subset this maps \emph{buy}~$\to$~1 and
\emph{strong buy}~$\to$~2.

\begin{figure}
\centering
\begin{tikzpicture}[
  font=\small,
  box/.style={draw, rounded corners=5pt, align=center,
              inner sep=6pt, line width=0.65pt},
  databox/.style={box, fill=blue!7, draw=blue!35,
                  minimum width=2.55cm, minimum height=1.45cm},
  agbox/.style={box, minimum width=2.55cm, minimum height=0.52cm,
                font=\footnotesize},
  Nbox/.style= {agbox, fill=blue!17,   draw=blue!55},
  Fbox/.style= {agbox, fill=orange!22, draw=orange!60},
  Dbox/.style= {agbox, fill=green!18,  draw=green!55},
  Mbox/.style= {agbox, fill=red!12,    draw=red!50},
  synthbox/.style={box, fill=violet!9, draw=violet!45,
                   minimum width=2.75cm, minimum height=1.45cm},
  embedbox/.style={box, fill=gray!11, draw=gray!50,
                   minimum width=2.75cm, minimum height=0.92cm},
  nnlsbox/.style= {box, fill=teal!10, draw=teal!55,
                   minimum width=3.65cm, minimum height=0.92cm},
  arr/.style={-{Stealth[length=5pt,width=4pt]},
              line width=0.9pt, color=black!45},
  lbl/.style={font=\scriptsize\itshape, color=black!50, inner sep=1.5pt},
  stagebox/.style={draw=black!15, fill=black!2, rounded corners=7pt,
                   dashed, line width=0.5pt, inner sep=8pt},
]

\node[databox] (data) at (0,0)
  {\textbf{Market Data}\\[4pt]
   \scriptsize news $\cdot$ fundamentals\\[-2pt]
   \scriptsize dynamics $\cdot$ macro};

\node[Nbox,  right=1.35cm of data, yshift= 1.08cm] (news)  {\textbf{News} agent};
\node[Fbox,  right=1.35cm of data, yshift= 0.36cm] (fund)  {\textbf{Fundamentals} agent};
\node[Dbox,  right=1.35cm of data, yshift=-0.36cm] (dyn)   {\textbf{Dynamics} agent};
\node[Mbox,  right=1.35cm of data, yshift=-1.08cm] (macro) {\textbf{Macro} agent};

\node[synthbox, right=1.35cm of fund, yshift=-0.36cm] (synth)
  {\textbf{Synthesis Agent}\\[5pt]
   \scriptsize thesis\\[-3pt]
   \scriptsize $+$\,recommendation};

\foreach \ag in {news,fund,dyn,macro} { \draw[arr] (data) -- (\ag); }
\foreach \ag in {news,fund,dyn,macro} { \draw[arr] (\ag)  -- (synth); }

\node[embedbox, below=1.35cm of synth] (embed)
  {\textbf{Text Embedding}\\[3pt]
   \scriptsize\texttt{text-emb-3-small}};

\draw[arr] (synth) --
  node[right=3pt, lbl]{thesis text} (embed);

\node[nnlsbox, left=1.2cm of embed] (nnls)
  {\textbf{NNLS Attribution}\\[3pt]
   \scriptsize
   $\hat{\vw}_{i,d}=(\wh_N,\wh_F,\wh_D,\wh_M)^{\!\top}$};

\draw[arr] (embed) --
  node[above=2pt, lbl]{$\mathbf{t}_{i,d}\!\in\!\mathbb{R}^{1536}$} (nnls);

\begin{pgfonlayer}{background}
  \node[stagebox,
        fit=(data)(news)(fund)(dyn)(macro)(synth),
        label={[font=\scriptsize\scshape,
                text=black!40, yshift=-2pt]above left:Stage 1 — Generation}]
       (stage1) {};
  \node[stagebox,
        fit=(embed)(nnls),
        label={[font=\scriptsize\scshape,
                text=black!40, yshift=2pt]below left:Stage 2 — Attribution}]
       (stage2) {};
\end{pgfonlayer}

\end{tikzpicture}
\caption{%
  \msai{} pipeline.
  \textit{Stage~1 (Generation):} four specialist agents independently analyse market data and produce focused text analyses; the synthesis agent reads all four summaries and generates a free-text equity thesis together with an ordinal recommendation---its explicit sentiment assessment of the stock.
  \textit{Stage~2 (Attribution):} the thesis text is encoded with
  \texttt{text-embedding-3-small} into a vector
  $\mathbf{t}_{i,d}\!\in\!\mathbb{R}^{1536}$; NNLS attribution then
  decomposes this embedding onto the four agent embeddings, recovering
  per-stock-date contribution weights
  $\hat{\vw}_{i,d}=(\wh_N,\wh_F,\wh_D,\wh_M)^\top$.}
\label{fig:pipeline}
\end{figure}
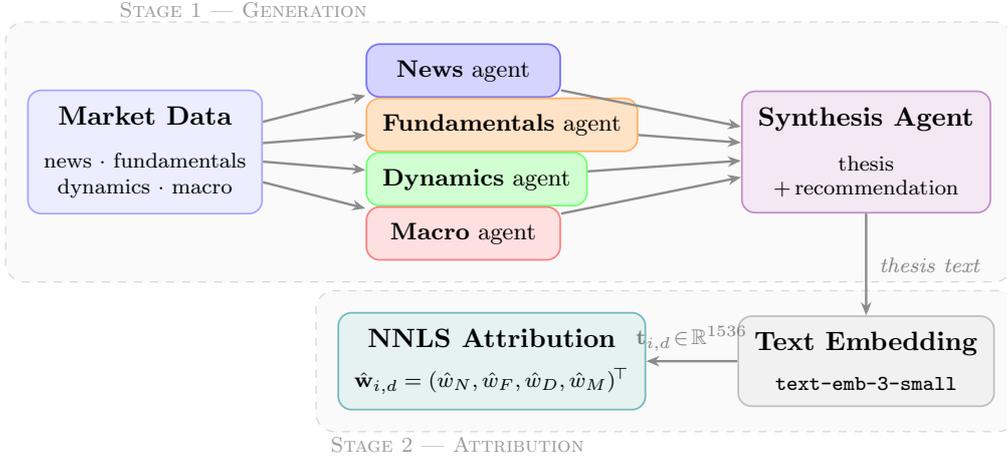

\paragraph{Pipeline:}
Figure~\ref{fig:pipeline} illustrates the architecture.
All text outputs---agent summaries and the synthesised thesis---are encoded with OpenAI's \texttt{text-embedding-3-small} ($D=1536$)  model without any post-processing.

\paragraph{Live generation and data integrity:}
All agent outputs are produced through live execution of \msai{} at each
observation date, not generated retroactively.  For observation dates beyond
the training cut-off of the underlying LLMs---which covers the entirety of
the \spx{} cohort (Sep 2024 onward) and the large majority of the \spone{}
cohort---news, earnings, and price data are necessarily absent from the
models' pre-training corpus, eliminating knowledge leakage as a potential
source of apparent outperformance.  This live-generation design is the
primary safeguard against look-ahead bias and is also why the empirical
analysis is anchored to these two specific fixed cohorts and observation
dates.

\paragraph{Fixed-cohort design:}
To avoid survivorship and look-ahead biases we use two fixed cohorts
(Table~\ref{tab:cohorts}).  The \spx{} cohort comprises the 467 stocks
present in the index from the post-expansion date onward; the \spone{}
cohort comprises 94 stocks over a longer horizon.  Observation dates follow
a first-Friday-of-the-month cadence; forward returns are one-month
buy-and-hold returns.

\begin{table}
\centering
\caption{Fixed-cohort design.}
\label{tab:cohorts}
\begin{tabular}{lcccc}
\toprule
Cohort & Stocks & Dates & Obs.\ rows & Period \\
\midrule
\spx{}  & 467 & 19 & 8{,}873 & Sep 2024 -- Mar 2026 \\
\spone{} & 94  & 35 & 3{,}290 & May 2023 -- Mar 2026 \\
\bottomrule
\end{tabular}
\end{table}

\section{Methodology}
\label{sec:methodology}

\subsection{Monte Carlo Portfolio Test}
\label{sec:mc_method}

Let $\mathcal{U}_d$ denote the full universe of stocks on date $d$ and let
$\mathcal{S}_d\subseteq\mathcal{U}_d$ be the set of stocks assigned
\emph{strong buy} on that date, with $|\mathcal{S}_d|=n_d$.  The actual
portfolio return on date $d$ is the equal-weight mean:
\[
  R^{\mathrm{SB}}_d = \frac{1}{n_d}\sum_{i\in\mathcal{S}_d} r_{i,d}^{(1\mathrm{m})},
\]
where $r_{i,d}^{(1\mathrm{m})}$ is the one-month forward return for stock $i$.

We also define the \emph{equal-weight (EW) universe benchmark} on date $d$
as the mean return of all stocks in $\mathcal{U}_d$:
\[
  R^{\mathrm{EW}}_d = \frac{1}{|\mathcal{U}_d|}\sum_{i\in\mathcal{U}_d}
    r_{i,d}^{(1\mathrm{m})}.
\]
This is the passive ``hold everything equally'' return for the covered
universe, approximating the RSP ETF (\spx{} cohort) and the EQWL ETF
(\spone{} cohort).  By linearity of expectation, $R^{\mathrm{EW}}_d$ equals
the expected value of any random equal-weight draw from $\mathcal{U}_d$,
so the EW benchmark and the MC null mean are mathematically equivalent;
any deviation in practice is Monte Carlo noise.

The Monte Carlo null distribution is constructed by, for each simulation
$k=1,\ldots,K$ ($K=10{,}000$), drawing $n_d$ stocks uniformly at random
\emph{without replacement} from $\mathcal{U}_d$ and computing their
equal-weight return $R^{(k)}_d$.  Sampling without replacement from the same
universe on the same date controls for universe composition, market timing,
and position count; the only variable across simulations is which stocks are
selected.

We aggregate across dates in two ways.  The \emph{mean-monthly} approach
takes the arithmetic average across dates:
\[
  \bar{R}^{\mathrm{SB}} = \frac{1}{T}\sum_{d=1}^T R^{\mathrm{SB}}_d,
  \qquad
  \bar{R}^{(k)} = \frac{1}{T}\sum_{d=1}^T R^{(k)}_d.
\]
The \emph{compound} approach takes the product of gross returns:
\[
  C^{\mathrm{SB}} = \prod_{d=1}^T (1+R^{\mathrm{SB}}_d) - 1,
  \qquad
  C^{(k)} = \prod_{d=1}^T (1+R^{(k)}_d) - 1.
\]
One-tailed empirical $p$-values are $p_{\mathrm{mean}}=K^{-1}\sum_k
\mathbf{1}[\bar{R}^{(k)}\ge\bar{R}^{\mathrm{SB}}]$ and analogously for
compound returns.  All results use $K=10{,}000$ and a fixed random seed for
reproducibility.

\subsection{NNLS Attribution}
\label{sec:nnls}

Let $\vt_{i,d}\in\RR^D$ be the thesis embedding for stock $i$ on date $d$,
and $\va^{(k)}_{i,d}\in\RR^D$ the embedding of agent $k$'s summary, where
$k\in\{\mathrm{news},\mathrm{fund},\mathrm{dyn},\mathrm{macro}\}$.
Form the agent matrix
$A_{i,d} = [\va^{(\mathrm{news})} \mid \va^{(\mathrm{fund})} \mid
\va^{(\mathrm{dyn})} \mid \va^{(\mathrm{macro})}] \in \RR^{D\times 4}$.

The NNLS problem
\begin{equation}
  \hat{\vw}_{i,d} = \arg\min_{\vw\ge\mathbf{0}}
    \bigl\|\vt_{i,d} - A_{i,d}\vw\bigr\|_2^2
  \label{eq:nnls}
\end{equation}
yields non-negative weights $\hat{\vw}_{i,d}=(\wh_N,\wh_F,\wh_D,\wh_M)^\top$
that represent each agent's contribution to the synthesis thesis.  Weights
are normalised to sum to one; if no feasible positive solution is found the
fallback is uniform weights.  Zero observations trigger this fallback in
either cohort.

\begin{figure}
\centering
\includegraphics[width=0.82\textwidth]{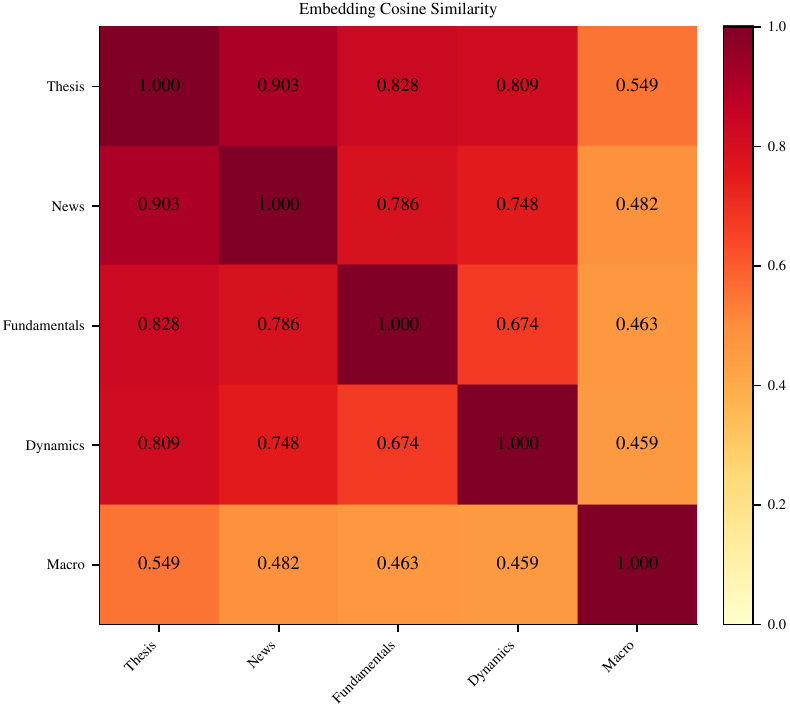}
\caption{%
  Cosine similarity heatmap (\spx{} cohort): thesis--agent cosines
  $C^{\mathrm{TA}}_k$ (top row), agent--agent cosines $C^{\mathrm{AA}}_{k,\ell}$
  (lower block), and mean thesis--reconstruction cosine
  $C^{\mathrm{TR}}=0.944$.  High pairwise agent cosines (0.46--0.79) motivate
  joint NNLS fitting over univariate cosine attribution.  The News agent's high
  thesis cosine (0.903) but near-zero pooled \IC{} ($+0.004$, buy+strong-buy universe)
  illustrates why cosine alone is an insufficient attribution signal.}
\label{fig:cosine}
\end{figure}

\paragraph{Why joint fitting rather than cosine alone:}
Since agent embeddings are mutually correlated (mean pairwise cosines
0.46--0.79; Figure~\ref{fig:cosine}), cosine similarity of the thesis with
a single agent conflates genuine agent influence with shared semantic
content.  NNLS jointly regresses the thesis onto all four agents
simultaneously, so the News agent's weight is reduced when Fundamentals
explains the same semantic mass more efficiently.  Figure~\ref{fig:cosine}
illustrates this concretely: the News agent has the highest mean thesis
cosine but a near-zero pooled \IC{}, while NNLS correctly reallocates
weight to Fundamentals and Macro, which carry positive \IC{}.

\paragraph{Attribution workflow:}
The NNLS analysis proceeds in two distinct stages.  In the
\emph{extraction} stage, Equation~\eqref{eq:nnls} is solved independently
for every stock--date pair, yielding a panel of normalised agent
contribution weights $\hat{w}_{k,i,d}$ that represent how much each agent
shaped the synthesis thesis.  In the \emph{validation} stage, these
extracted weights are treated as continuous signals and evaluated against
one-month forward returns via Spearman rank correlation---both pooled
across all stock--date pairs (pooled \IC{}) and cross-sectionally for each
date (date-level \IC{} and \ICIR{}).  This two-stage structure cleanly
separates the unsupervised embedding decomposition from the downstream
test of whether the recovered weights carry predictive return information,
ensuring that the attribution and evaluation steps are methodologically
independent.

\subsection{Cosine Diagnostics}
\label{sec:cosine}

Three cosine families provide diagnostic checks on the NNLS attribution.
\emph{Thesis--agent cosines} $C^{\mathrm{TA}}_{k}(i,d)=
\cos(\vt_{i,d},\va^{(k)}_{i,d})$ measure univariate semantic proximity.
\emph{Agent--agent cosines} $C^{\mathrm{AA}}_{k,\ell}(i,d)$ quantify
collinearity, motivating joint fitting.
\emph{Thesis--reconstruction cosines}
$C^{\mathrm{TR}}_{i,d}=\cos(\vt_{i,d},\hat{\vt}_{i,d})$ assess how well
the four-agent subspace spans the thesis embedding, where
$\hat{\vt}_{i,d}=A_{i,d}\hat{\vw}_{i,d}$.

Spearman rank correlation between $C^{\mathrm{TA}}_{k}$ and $\wh_k$ across
the panel validates that NNLS weights agree directionally with cosine
rankings while correcting for collinearity.

\subsection{Information Coefficient and \ICIR{}}
\label{sec:ic}

The \emph{pooled \IC{}} is the Spearman rank correlation between a signal
$x_{i,d}$ and the one-month forward return $r_{i,d}^{(1\mathrm{m})}$,
computed over all $N$ stock--date pairs.  This treats each observation as
independent.  However, stocks within a date share common factor exposures
(market beta, sector), so returns within a date are cross-sectionally
correlated.  Pooled \IC{} therefore conflates stock-level fixed effects
with genuine timing signal; we use it directionally only.

The \emph{date-level \IC{}} computes one cross-sectional Spearman
correlation per date $d$ using all stocks in the cohort.  The
$T$ date-level ICs are approximately independent (non-overlapping monthly
return windows) and are tested with a one-sample $t$-test:
\begin{equation}
  \overline{\IC} = \frac{1}{T}\sum_{d=1}^T \IC_d,
  \qquad
  \ICIR{} = \frac{\overline{\IC}}{\mathrm{std}(\IC_d)},
  \qquad
  t = \ICIR{}\cdot\sqrt{T}.
\end{equation}
With $T=19$ (\spx{}) the threshold for $p=0.05$ is $|\ICIR{}|>0.47$;
with $T=35$ (\spone{}) it is $|\ICIR{}|>0.34$.

\paragraph{Actionable-universe scope:}
All \IC{} analysis in this paper is restricted to the \textit{buy and
strong-buy observations only}: the two signal classes for which \msai{}
recommends a long position.  Hold, sell, and strong-sell observations are
excluded from every \IC{} computation.  The rationale is direct: for
non-actionable observations no portfolio position is taken, so testing
whether agent weights or the ordinal score rank those stocks is irrelevant
to portfolio performance.  This scope fully aligns the \IC{} test with the
Monte Carlo test, which also operates on the actionable universe.  The
ordinal score (defined in Section~\ref{sec:system}) is the explicit signal
tested against forward returns in the date-level \IC{} analysis.  The
average per-date sample size in this subset is $\sim$74 stocks (\spx{})
and $\sim$19 stocks (\spone{}).

\section{Data and Descriptive Statistics}
\label{sec:data}

The two fixed cohorts introduced in Section~\ref{sec:system} form the basis
of all empirical analysis.  Before presenting performance results we
characterise the composition of the signal panel across both cohorts,
establishing baseline distributional facts that contextualise the
portfolio-level and attribution findings.

\paragraph{Signal class distribution:}
Table~\ref{tab:signal_dist} shows the distribution of ordinal
recommendations across both cohorts.  Hold dominates ($\sim$79\% in
\spx{}, $\sim$76\% in \spone{}), with strong-buy constituting 7.5\% and
10.5\% respectively.  The relative scarcity of strong-buy signals is
important context for the Monte Carlo test: the system concentrates its
highest-conviction calls on a small fraction of the universe, and it is
precisely this concentrated selection that the null distribution tests.

\begin{table}
\centering
\caption{Signal class distribution across both cohorts.}
\label{tab:signal_dist}
\begin{tabular}{lrrrr}
\toprule
Signal class & \multicolumn{2}{c}{\spx{}} & \multicolumn{2}{c}{\spone{}} \\
\cmidrule(lr){2-3}\cmidrule(lr){4-5}
 & Count & \% & Count & \% \\
\midrule
Strong sell & 180  &  2.0 & 48  &  1.5 \\
Sell        & 286  &  3.2 & 77  &  2.3 \\
Hold        & 6{,}992 & 78.8 & 2{,}508 & 76.2 \\
Buy         & 749  &  8.4 & 310 &  9.4 \\
Strong buy  & 666  &  7.5 & 347 & 10.5 \\
\midrule
Total       & 8{,}873 & 100 & 3{,}290 & 100 \\
\bottomrule
\end{tabular}
\end{table}

\begin{figure}
\centering
\includegraphics[width=\textwidth]{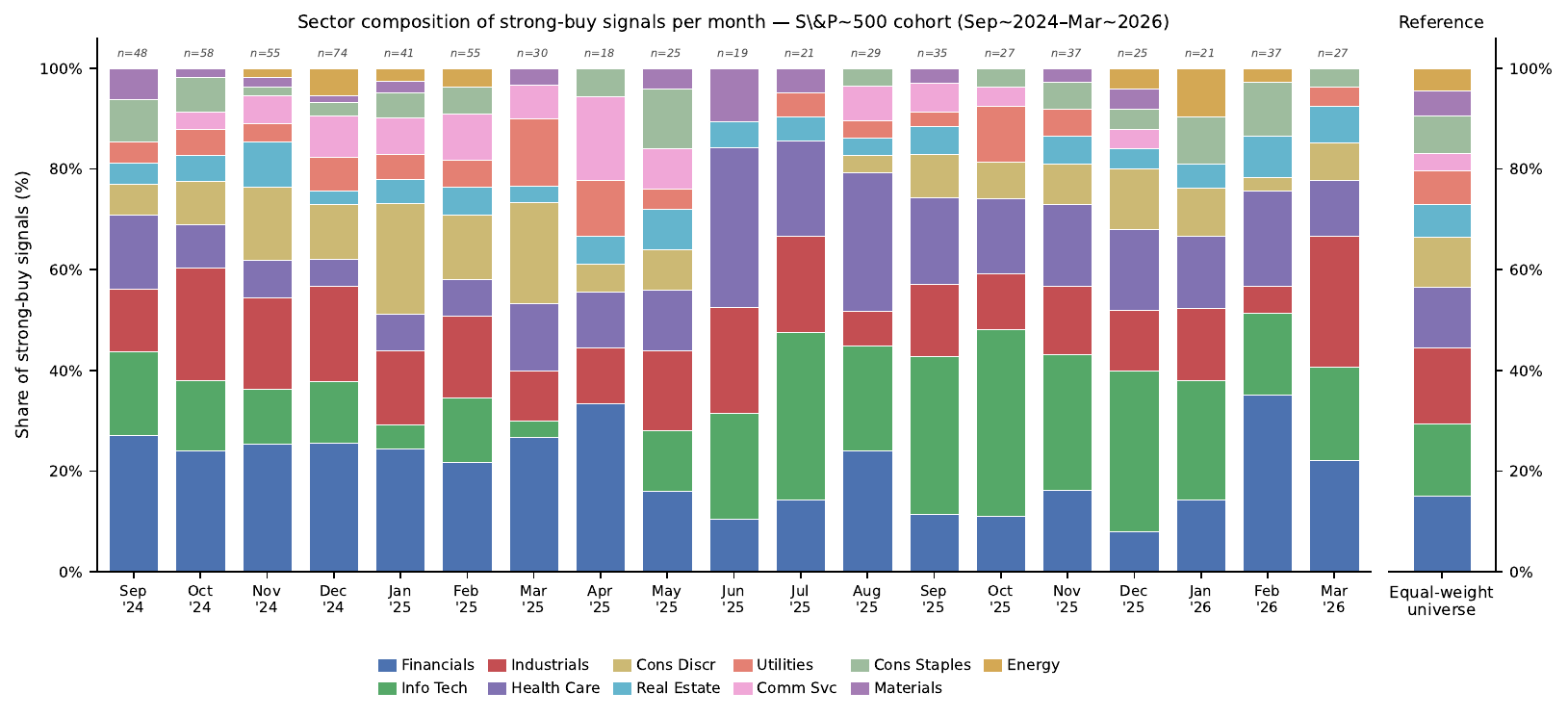}
\caption{%
  Sector composition of the equal-weight strong-buy basket per month
  (\spx{} cohort, Sep~2024--Mar~2026) alongside the equal-weight
  universe as a reference (rightmost bar).  Each stacked bar sums to
  100\%; $n$ labels show the number of strong-buy picks that month.
  Financials dominate the early period (Sep~2024--Feb~2025;
  $\sim$24.8\% share vs 15.1\% in the universe), while Information
  Technology becomes the largest sector from Mar~2025 onward
  ($\sim$21.7\%).  Energy (1.8\% vs 4.4\%), Materials (2.2\% vs 5.1\%),
  and Consumer Staples (4.7\% vs 7.4\%) are persistently
  under-represented.}
\label{fig:sector_dist}
\end{figure}

\paragraph{Sector distribution of strong-buy signals:}
Strong-buy signals are not uniformly distributed across the eleven
\spx{} GICS sectors, nor is the sector composition stable over time.
Figure~\ref{fig:sector_dist} shows the sector share of each monthly
strong-buy basket (equal-weight; one bar per observation date) alongside
the equal-weight sector composition of the full universe as a reference.
Pooled over the full 19-month period, Financials account for 21.8\% of
strong-buy picks, $+6.7$ percentage points above their 15.1\% universe
weight. Information Technology (17.0\% vs 14.2\% in the universe) is modestly
over-represented in aggregate, while Energy (1.8\% vs 4.4\%),
Materials (2.2\% vs 5.1\%), and Consumer Staples (4.7\% vs 7.4\%) are
persistently under-represented. The temporal pattern demonstrates the sector rotation over the period and confirms that the strong-buy selection is not a static bet on a single sector (see  Section~\ref{sec:sector}).

\section{Results}
\label{sec:results}

\subsection{Monte Carlo: Strong-Buy Portfolio vs Random Selection}
\label{sec:mc_results}

\begin{figure}
\centering
\includegraphics[width=\textwidth]{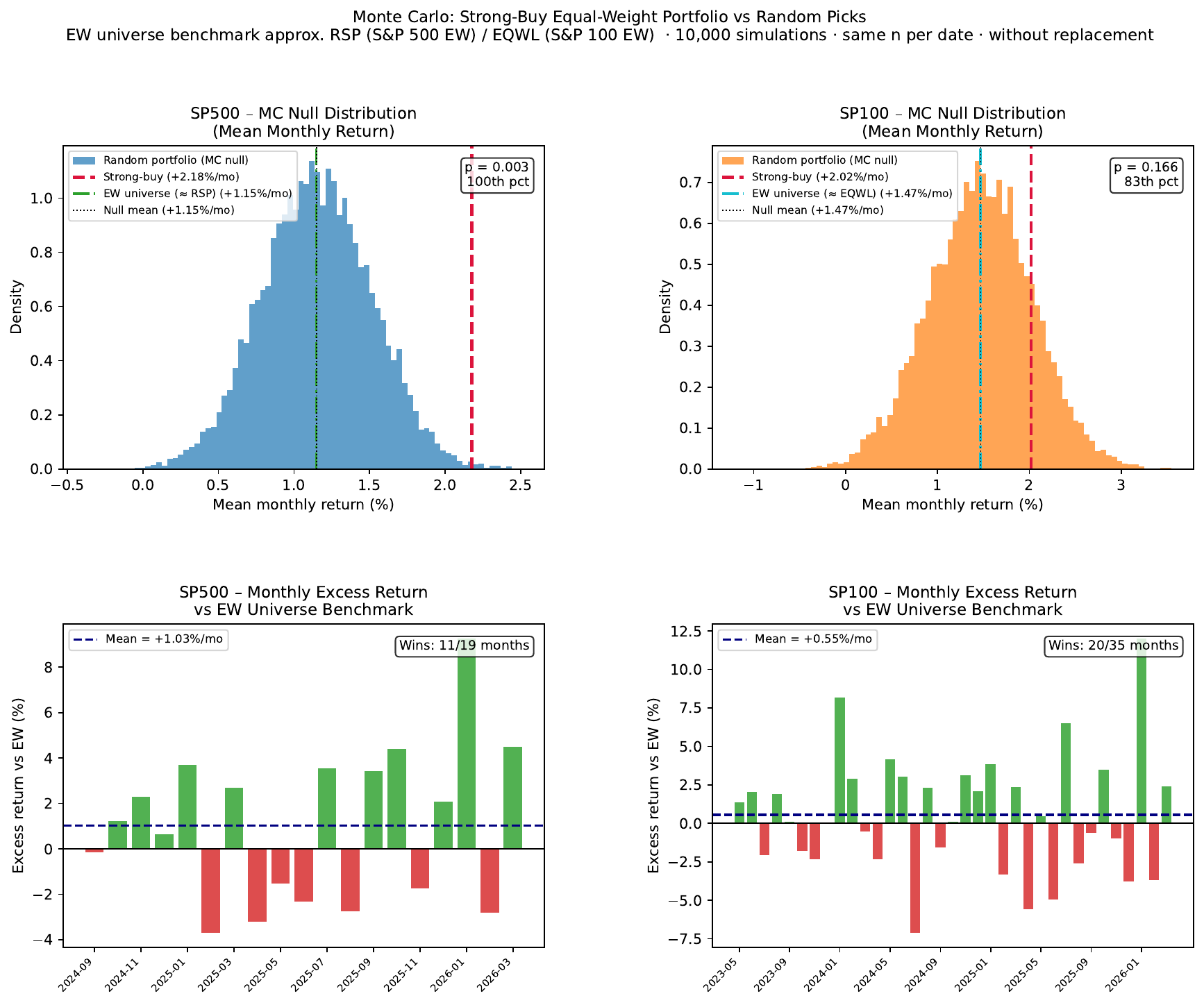}
\caption{%
  Monte Carlo null distributions of mean monthly equal-weight portfolio
  returns (10{,}000 simulations).  Top panels: \spx{} (left) and \spone{}
  (right).  Red dashed line: actual strong-buy portfolio return; green
  dash-dot line: equal-weight universe benchmark ($\approx$\,RSP / EQWL).
  Black dotted: MC null mean, which is mathematically equivalent to the EW
  benchmark (see Section~\ref{sec:mc_method}).  Bottom panels:
  month-by-month excess return of the strong-buy portfolio relative to the
  EW benchmark; green bars indicate outperforming months.  The \spx{} result
  sits at the 99.7th MC percentile ($p=0.003$); \spone{} at the 83rd
  ($p=0.17$).}
\label{fig:mc_null}
\end{figure}

\begin{figure}
\centering
\includegraphics[width=\textwidth]{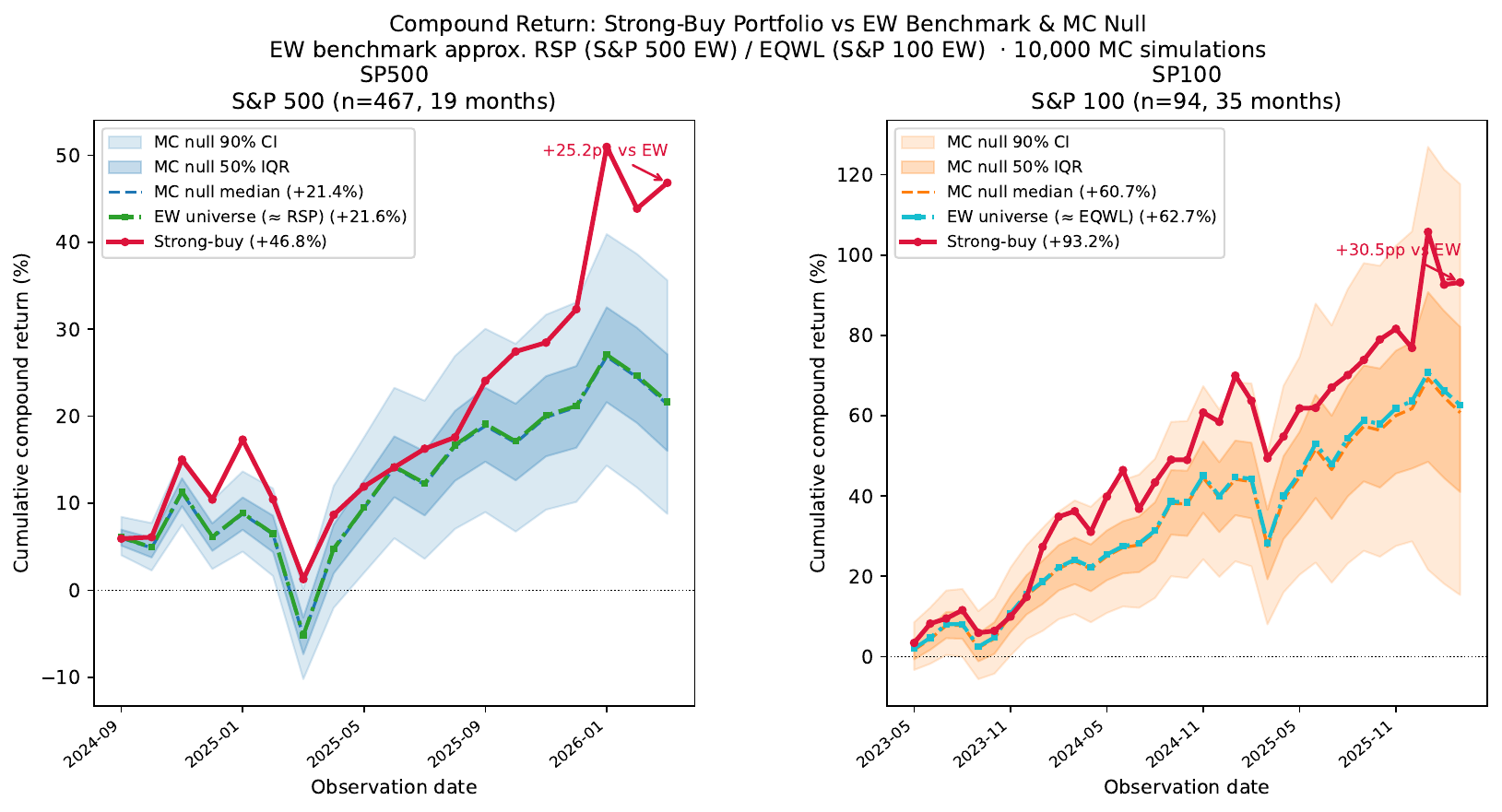}
\caption{%
  Compound growth of the strong-buy equal-weight portfolio (red solid line)
  versus the equal-weight universe benchmark (green dash-dot; $\approx$\,RSP
  / EQWL) and the Monte Carlo null (cohort-coloured dashed: median; shaded
  bands: 50\% IQR and 90\% CI from 10{,}000 simulations).  Left: \spx{};
  right: \spone{}.  Annotations show final excess over the EW benchmark.
  The EW benchmark and MC null median are nearly identical, confirming the
  null captures passive EW performance.  The \spx{} strong-buy trajectory
  exceeds the 90th-percentile null boundary for the majority of the sample.}
\label{fig:mc_compound}
\end{figure}

Table~\ref{tab:mc} and Figures~\ref{fig:mc_null}--\ref{fig:mc_compound}
present the Monte Carlo results.

\begin{table}
\centering
\caption{Monte Carlo results: strong-buy equal-weight portfolio vs (i) a
  passive equal-weight benchmark of all covered stocks (EW universe,
  approximating RSP for \spx{} and EQWL for \spone{}) and (ii) 10{,}000
  random same-sized portfolios from the same universe.
  $p$-values are one-tailed empirical (fraction of MC simulations $\ge$ actual).}
\label{tab:mc}
\begin{tabular}{lcc}
\toprule
Metric & \spx{} (19 dates) & \spone{} (35 dates) \\
\midrule
Avg.\ strong-buy picks / month & 35.1 & 9.9 \\
\addlinespace
\multicolumn{3}{l}{\textit{Mean-monthly approach}} \\
Strong-buy mean monthly return     & $+2.18\%$ & $+2.02\%$ \\
EW benchmark (RSP / EQWL)         & $+1.15\%$ & $+1.47\%$ \\
MC null median monthly return      & $+1.15\%$ & $+1.47\%$ \\
\textbf{Excess vs EW benchmark}    & $\mathbf{+1.02\%}$ & $+0.55\%$ \\
Percentile rank in MC null         & \textbf{99.7th} & 83.4th \\
$p$-value (vs MC null)             & $\mathbf{0.003}$ & 0.166 \\
\addlinespace
\multicolumn{3}{l}{\textit{Compound return (full period)}} \\
Strong-buy compound return         & $\mathbf{+46.8\%}$ & $+93.2\%$ \\
EW benchmark compound return       & $+21.6\%$ & $+62.7\%$ \\
MC null median compound return     & $+21.4\%$ & $+60.7\%$ \\
\textbf{Excess vs EW benchmark}    & $\mathbf{+25.2\,\text{pp}}$ & $+30.5\,\text{pp}$ \\
$p$-value (vs MC null)             & $\mathbf{0.003}$ & 0.163 \\
\addlinespace
\multicolumn{3}{l}{\textit{Win rate}} \\
Months where SB $>$ EW benchmark   & 11/19 (57.9\%) & 20/35 (57.1\%) \\
\bottomrule
\end{tabular}
\end{table}

\paragraph{\spx{} cohort (primary result):}
The strong-buy portfolio earns $+2.18\%$/month versus the EW universe
benchmark of $+1.15\%$ (approximating RSP), an excess of $+1.02\%$/month.
The result sits at the 99.7th percentile of 10{,}000 Monte Carlo alternatives
($p=0.003$).  The compound return of $+46.8\%$ over 19 months is $+25.2$
percentage points above the EW benchmark of $+21.6\%$, and the MC null
median is $+21.4\%$ ($p=0.003$).  Strong-buy picks beat the EW benchmark in
11 of 19 months (57.9\%).

Figure~\ref{fig:mc_null} shows the null distribution of mean monthly
returns for both cohorts.  The actual \spx{} strong-buy return (red dashed
line) sits deep in the right tail, clearly separated from the bulk of
the null.

\paragraph{\spone{} cohort (robustness):}
The directional pattern holds: $+0.55\%$/month above the EW benchmark of
$+1.47\%$ (approximating EQWL), 83rd MC percentile, compound return
$+93.2\%$ versus the EW benchmark of $+62.7\%$ ($+30.5$ percentage points)
and the MC null median of $+60.7\%$.  Formal significance is not reached
($p=0.17$).  The wider null distribution reflects the small average selection
of $\sim\!10$ stocks per date: drawing 10 stocks at random from 94 produces
high variance in monthly returns, reducing statistical power.  The result is
consistent with the \spx{} finding but does not independently confirm it.

Figure~\ref{fig:mc_compound} plots compound growth paths for both cohorts.
The actual strong-buy trajectory lies above the null median throughout most
of the \spx{} period and comfortably above for \spone{}, though the
\spone{} CI is wide.

\paragraph{Interpretation:}
The EW universe benchmark provides a concrete passive reference: buying
all covered stocks with equal weight on every date (approximating RSP for
\spx{} and EQWL for \spone{}).  The Monte Carlo null formalises this
further by testing whether the strong-buy \emph{selection} adds value
beyond passive EW; because the MC null's expected value equals the EW
benchmark, the two benchmarks are equivalent in expectation and both
confirm the same excess return.  Market-level and sector-level returns
cancel in the comparison since both actual and simulated portfolios are
drawn from the same universe on the same dates with the same equal
weighting.  The \spx{} result is robust across both aggregation approaches
(mean-monthly and compound) with the same $p$-value ($0.003$).  To
contextualise the magnitude, the $+1.02\%$/month gross excess return
corresponds to an annualised alpha of approximately $+12.4\%$.  At a
monthly rebalancing frequency across ${\sim}35$ large-cap equal-weight
positions, typical implementation drag (bid-ask spread, market impact)
would be well below 30 bps/month, leaving substantial headroom before the
signal becomes unprofitable net of costs.

Note that the MC test and the \IC{} analysis address different questions.
IC measures whether the agent weights rank individual stocks correctly in
the cross-section on a given date.  The MC test measures whether the subset
of stocks selected as strong-buy outperforms an equally-sized random subset.
Both can be informative simultaneously: the IC can be near-zero because most
of the cross-section is holds (rank correlation diluted), while the extreme
tail of strong-buys still earns a meaningful average return premium.

\subsection{NNLS Reconstruction Quality and Cosine Validation}
\label{sec:attribution}

Having established portfolio-level selection skill, we turn to NNLS
attribution to characterise how the four agents collectively shape the
thesis and whether their contributions vary across sectors and regimes.

\paragraph{Reconstruction quality:}
The four-agent subspace reconstructs thesis embeddings with high fidelity:
mean $C^{\mathrm{TR}}=0.944$ (\spx{}) and $0.936$ (\spone{}).  These figures
confirm that the NNLS problem is well-posed and that the agent summaries
collectively span nearly all semantic variance in the synthesis thesis.

\paragraph{Cosine--weight agreement:}
Table~\ref{tab:cosine} reports the Spearman correlation between
$C^{\mathrm{TA}}_k$ and $\hat{w}_k$ across the panel.  All correlations are
strongly positive (up to $\rho_s=0.826$, all $p<10^{-85}$), confirming
that NNLS weights agree directionally with cosine rankings while correcting
for inter-agent collinearity.

\begin{table}
\centering
\caption{Spearman $\rho_s$ between thesis--agent cosine $C^{\mathrm{TA}}_k$
  and normalised NNLS weight $\hat{w}_k$, across all stock--date pairs.
  All $p<10^{-85}$.}
\label{tab:cosine}
\begin{tabular}{lcccc}
\toprule
 & \multicolumn{2}{c}{\spx{}} & \multicolumn{2}{c}{\spone{}} \\
\cmidrule(lr){2-3}\cmidrule(lr){4-5}
Agent & $\rho_s$ & $n$ & $\rho_s$ & $n$ \\
\midrule
News         & 0.571 & 8{,}873 & 0.534 & 3{,}290 \\
Fundamentals & 0.826 & 8{,}873 & 0.795 & 3{,}290 \\
Dynamics     & 0.724 & 8{,}873 & 0.701 & 3{,}290 \\
Macro        & 0.648 & 8{,}873 & 0.631 & 3{,}290 \\
\addlinespace
Thesis reconstruction ($C^{\mathrm{TR}}$, mean) & $0.944$ & -- & $0.936$ & -- \\
\bottomrule
\end{tabular}
\end{table}

\subsection{Downside Behaviour of Long Signals}
\label{sec:risk}

Table~\ref{tab:risk} reports the equal-weight mean return for positive-return
months (upside), negative-return months (downside), hit rate, and the
upside/downside ratio, broken down by signal class.  A bootstrap confidence
interval on the ratio difference $\Delta\mathrm{UpDn} =
\mathrm{UpDn}_{\mathrm{strong buy}} - \mathrm{UpDn}_{\mathrm{hold}}$
is computed from $5{,}000$ resamples.

\begin{table}
\centering
\caption{Signal risk profile (\spx{} cohort, 1-month horizon).
  Upside (Upside$_+$): mean return on months where signal class has positive
  returns.  Downside (Down$_-$): mean return on negative months (shown as
  negative).  Hit rate: fraction of months with positive return.
  UpDn: $|\bar r^+/\bar r^-|$.  Bootstrap CI on
  $\Delta$UpDn (strong buy $-$ hold) from 5{,}000 resamples.}
\label{tab:risk}
\begin{tabular}{lcccccc}
\toprule
Signal & Upside$_+$ & Down$_-$ & Hit\% & UpDn & $\Delta$UpDn & 95\% CI \\
\midrule
Strong buy & $+6.39\%$ & $-5.95\%$ & 58.4 & 1.07 & $+0.17$ & $[-0.03,\,+0.37]$ \\
Buy        & $+5.93\%$ & $-6.00\%$ & 52.1 & 0.99 & $+0.09$ & $[-0.11,\,+0.28]$ \\
Hold       & $+5.27\%$ & $-6.71\%$ & 53.4 & 0.79 & ref.    & -- \\
\bottomrule
\end{tabular}
\end{table}

Strong-buy stocks show a smaller mean loss on negative months
($-5.95\%$ vs $-6.71\%$ for hold) without a corresponding reduction in
upside ($+6.39\%$ vs $+5.27\%$).  The bootstrap $\Delta\mathrm{UpDn}$
confidence interval is $[-0.03,+0.37]$ with one-tailed $p=0.050$---
directionally positive but spanning zero, so this result should be read as a
descriptional regularity from the available data rather than a formally
established effect.

\begin{figure}
\centering
\includegraphics[width=\textwidth]{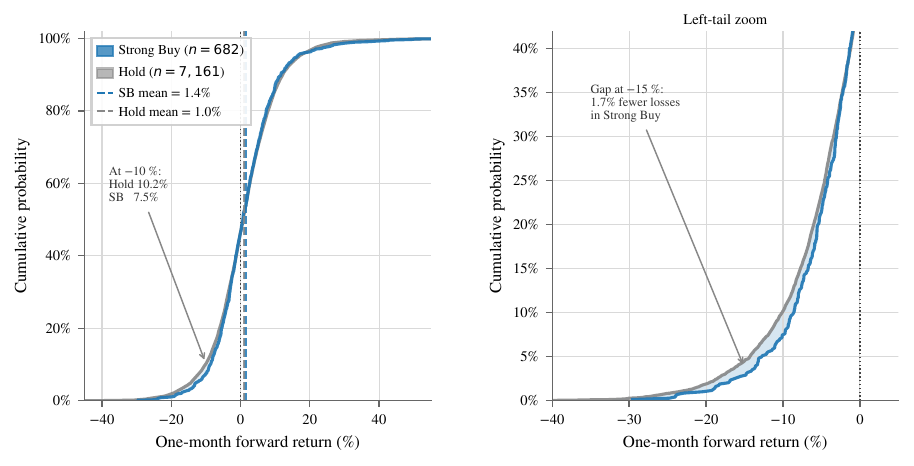}
\caption{%
  Empirical CDF comparison of one-month forward returns for strong-buy
  (blue) and hold (grey) signal classes (\spx{} cohort, $n=682$ and
  $n=7{,}161$ respectively).
  \textit{Left panel:} full CDF; dashed vertical lines mark the
  respective means; the shaded region highlights the left-tail gap
  (returns $<0$) where the two curves diverge most.
  \textit{Right panel:} left-tail zoom (returns $\leq+5\%$); the blue
  shading quantifies the probability mass by which strong-buy
  undercuts hold at each loss level.  At $-10\%$ the strong-buy
  exceedance probability is $7.5\%$ vs $10.2\%$ for hold---a
  $2.7$\,pp reduction in left-tail risk consistent with the
  upside/downside ratio in Table~\ref{tab:risk}.  The right tails
  converge, confirming the edge is concentrated in downside
  protection rather than higher return magnitudes.}
\label{fig:return_hist}
\end{figure}

Figure~\ref{fig:return_hist} uses empirical CDFs to compare the return
distributions of the strong-buy and hold signal classes.  The CDF is the
natural tool for this comparison: at any loss threshold on the left tail,
the strong-buy curve lies \emph{below} the hold curve, meaning a strictly
lower probability of incurring losses of that magnitude.  The left-tail
zoom (right panel) makes this gap concrete---at $-10\%$, strong-buy has a
$7.5\%$ exceedance probability versus $10.2\%$ for hold---confirming that
the upside/downside asymmetry in Table~\ref{tab:risk} is driven by
truncation of extreme negative returns rather than by elevated return
magnitudes at the right tail, where the two distributions converge.

The Mann--Whitney one-sided test comparing $|r|$ distributions of long
signals vs hold is not significant ($p=0.93$ for \spx{}, $p=0.60$ for
\spone{}), confirming that the edge is not driven by higher return
magnitudes among long signals.

\textbf{Sell-side note:} Sell and strong-sell stocks show positive mean one-month returns
($+1.65\%$ and $+2.98\%$ respectively for \spx{}), indicating contrarian
bearish calls over this period.  One plausible contributing mechanism is
that fundamentally weak stocks---precisely those flagged bearishly by the
system---are often prone to short squeezes in risk-on environments, where
high short interest and forced covering can generate strong short-horizon
returns that are independent of underlying quality.  The sample is too
small ($n<470$ total sell/strong-sell observations) for formal testing;
this pattern is flagged but not treated as an established result.

\subsection{Agent Weights Carry Forward-Return Information}
\label{sec:ic_results}

All \IC{} analysis in this section is restricted to the
\textit{buy and strong-buy universe} (Section~\ref{sec:ic}): observations
where \msai{} issues an actionable bullish signal.

Mean one-month forward returns are monotone in both cohorts:
buy $+1.35\%$ $<$ strong-buy $+1.47\%$ (\spx{}); and buy $+1.39\%$ $<$
strong-buy $+1.67\%$ (\spone{}), confirming that the ordinal label carries a
directional return signal within the actionable subset.

\begin{figure}
\centering
\includegraphics[width=\textwidth]{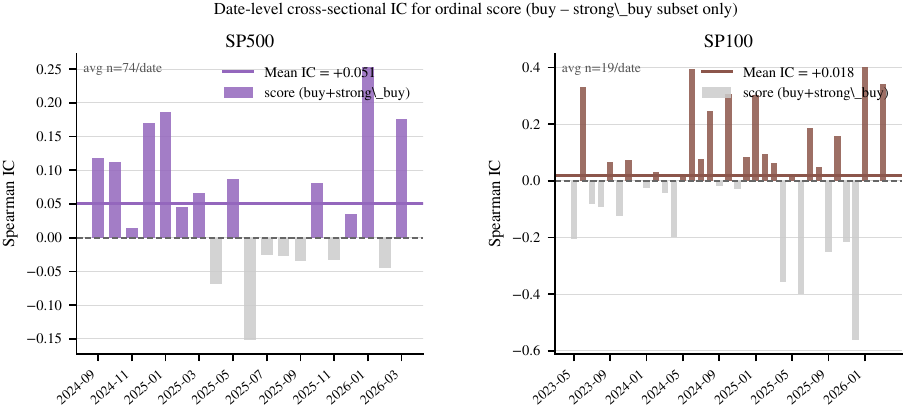}
\caption{%
  Date-level cross-sectional \IC{} for the ordinal score computed on the
  \textit{buy and strong-buy subset} (neutral/hold, sell, and strong-sell
  observations excluded).  Left panel: \spx{} (19 dates, avg 74 stocks/date);
  right panel: \spone{} (35 dates, avg 19 stocks/date).  Bars show per-date
  IC; bars shaded in colour indicate positive-IC dates; the horizontal line
  marks the date-level mean (\spx{}: $+0.051$, \spone{}: $+0.018$).
  The \spx{} \ICIR{} of $+0.489$ exceeds the $p=0.05$ significance
  threshold ($|\ICIR{}|>0.47$ at $T=19$), yielding $p=0.024$ ($t=+2.13$).}
\label{fig:ic_date}
\end{figure}

Figure~\ref{fig:ic_date} plots date-level \IC{} for the ordinal score
in both cohorts.  Table~\ref{tab:agent_ic} reports pooled \IC{},
date-level mean \IC{}, \ICIR{}, and the date-level $t$-test for all agents
and the score, all computed in the buy+strong-buy universe.

\begin{table}
\centering
\caption{Pooled and date-level \IC{} for agent NNLS weights and the ordinal
  score, computed on the \textit{buy and strong-buy universe only}
  (hold, sell, and strong-sell excluded; avg 74 stocks/date on \spx{},
  19 on \spone{}).  Pooled \IC{}: Spearman over all stock--date pairs
  (conflates stock fixed effects with timing; directional).
  Date-level \IC{}: mean of $T$ per-date cross-sectional correlations;
  $t$-test against zero; \ICIR{} = mean/std.
  Bold: entries with $p<0.05$.}
\label{tab:agent_ic}

\smallskip
\textit{Panel A: \spx{} cohort ($T=19$ dates)}
\begin{tabular}{lcccccc}
\toprule
Agent / Signal & Pool \IC{} & Pool $p$ & DL mean \IC{} & \ICIR{} & $t$ & $p_t$ \\
\midrule
News         & $+0.004$ & 0.887 & $-0.035$ & $-0.327$ & $-1.42$ & 0.086 \\
Fundamentals & $\mathbf{+0.052}$ & \textbf{0.049} & $+0.012$ & $+0.092$ & $+0.40$ & 0.346 \\
Dynamics     & $\mathbf{-0.069}$ & \textbf{0.009} & $+0.019$ & $+0.135$ & $+0.59$ & 0.282 \\
Macro        & $+0.030$ & 0.257 & $+0.016$ & $+0.113$ & $+0.49$ & 0.314 \\
\midrule
Score        & $+0.006$ & 0.822 & $\mathbf{+0.051}$ & $\mathbf{+0.489}$ & $\mathbf{+2.13}$ & \textbf{0.024} \\
\bottomrule
\end{tabular}

\smallskip
\textit{Panel B: \spone{} cohort ($T=35$ dates)}
\begin{tabular}{lcccccc}
\toprule
Agent / Signal & Pool \IC{} & Pool $p$ & DL mean \IC{} & \ICIR{} & $t$ & $p_t$ \\
\midrule
News         & $+0.016$ & 0.688 & $+0.047$ & $+0.189$ & $+1.12$ & 0.135 \\
Fundamentals & $-0.025$ & 0.529 & $-0.062$ & $-0.218$ & $-1.29$ & 0.103 \\
Dynamics     & $-0.040$ & 0.311 & $-0.035$ & $-0.150$ & $-0.89$ & 0.191 \\
Macro        & $\mathbf{+0.079}$ & \textbf{0.042} & $+0.033$ & $+0.138$ & $+0.82$ & 0.210 \\
\midrule
Score        & $+0.013$ & 0.749 & $+0.018$ & $+0.080$ & $+0.48$ & 0.319 \\
\bottomrule
\end{tabular}
\end{table}

\begin{figure}
\centering
\includegraphics[width=\textwidth]{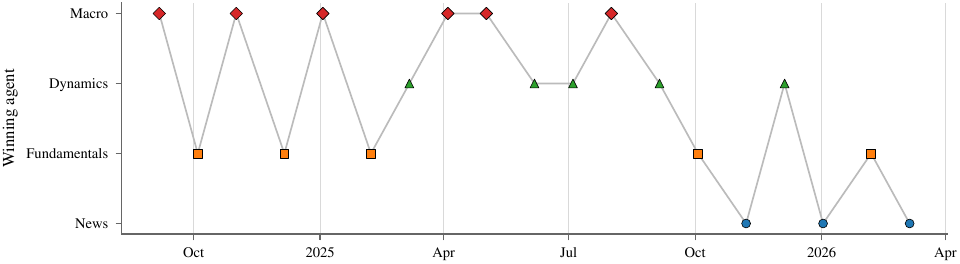}
\caption{%
  Best-agent timeline (\spx{} cohort, buy\,+\,strong-buy universe only):
  which agent achieves the highest cross-sectional \IC{} on each of the 19
  observation dates.  Macro leads on 6 dates, Dynamics on 5, Fundamentals
  on 5, and News on 3.  The rotation of the dominant agent is direct
  evidence for the adaptive integration hypothesis: no single agent
  persistently determines thesis quality, and the agent that leads changes
  with sector conditions and market regime.  Macro's six leading dates
  cluster around distinct macro-driven episodes
  (Table~\ref{tab:macro_episodes}).}
\label{fig:best_agent}
\end{figure}

\paragraph{Pooled \IC{}:}
Within the buy+strong-buy universe, Fundamentals ($+0.052$, $p=0.049$)
carries the largest positive pooled \IC{} on \spx{}; Macro dominates on
\spone{} ($+0.079$, $p=0.042$), reflecting the greater relevance of
macroeconomic context in a concentrated 94-stock universe.  Dynamics
shows a significantly negative pooled \IC{} on \spx{} ($-0.069$,
$p=0.009$) yet leads as the best agent on 5 of 19 dates
(Figure~\ref{fig:best_agent}), revealing a regime-conditional role:
when price-action signals are aligned with the broader trend the
synthesis elevates Dynamics weight and it contributes positively; in
aggregate across all dates it is a drag, consistent with momentum being
an episodic rather than persistent predictor.  This divergence between
pooled and date-level behaviour illustrates why no single agent can be
evaluated in isolation.  As noted in Section~\ref{sec:ic}, pooled \IC{}
is partly driven by stock-level fixed effects and should be read
directionally.

Macro's six leading dates cluster around identifiable macro-driven
episodes in which economy-wide forces, rather than idiosyncratic stock
dynamics, were the primary driver of return differentiation
(Table~\ref{tab:macro_episodes}).  In each case the synthesis agent
increased Macro weight precisely when broad policy or regime shifts
dominated cross-sectional dispersion.

\begin{table}
\centering
\caption{Macro-agent leading dates (\spx{} cohort) and the coincident
  macro regime shift.}
\label{tab:macro_episodes}
\small
\begin{tabular}{l l p{7.8cm}}
\toprule
Date & Macro event & Cross-sectional effect \\
\midrule
Sep 2024 & First Fed rate cut & Rotation into rate-sensitive names; start of easing cycle \\
Nov 2024 & US presidential election & ``Trump trade'' dispersion: financials, energy, defence up, rate-sensitive down \\
Jan 2025 & Inauguration \& policy signalling & Extension of election-driven sectoral rotation \\
Apr 2025 & ``Liberation Day'' tariffs & Extreme dispersion by trade exposure \\
May 2025 & US--China tariff escalation & Continued trade-policy-driven rotation \\
Aug 2025 & Recession concerns \& Fed path & Macro-driven defensive rotation \\
\bottomrule
\end{tabular}
\end{table}

\paragraph{Date-level \IC{} and the score result:}
The ordinal score's date-level \IC{} on \spx{} reaches mean $+0.051$, $\ICIR{}=+0.489$, and is statistically significant at $p=0.024$
($t=+2.13$, $T=19$ dates), confirming that the recommendation label itself distinguishes buy from strong-buy returns within the actionable universe.  On \spone{}, with only $\sim\!19$ stocks per date, the score \ICIR{} is $+0.080$
($p=0.319$)---the test remains underpowered at this sample size.  Agent date-level \IC{}s are small and insignificant for all four agents in both
cohorts (all $p>0.08$), consistent with agent weights encoding
cross-sectional return information at the pooled level rather than via
consistent month-by-month timing.

\paragraph{Information compression:}
On \spx{}, the Fundamentals pooled \IC{} ($+0.052$) is $\sim$9$\times$ the
ordinal score's pooled \IC{} ($+0.006$).  On \spone{}, the Macro pooled
\IC{} ($+0.079$) is $\sim$6$\times$ the ordinal score's pooled \IC{}
($+0.013$).  Both measures are computed in the same buy+strong-buy universe,
so the comparison is valid.  The continuous agent-blending structure encodes
cross-sectional return information at a finer grain than the discrete recommendation label.

\paragraph{Interpretation:}
The IC and MC results are jointly consistent with the synthesis agent
acting as an adaptive integrator: drawing more heavily on Fundamentals
when stock-specific quality signals dominate, on Macro when sector or
market conditions are the primary driver, and on Dynamics selectively
when price-action momentum is informative.  Agent weights behave as
persistent, factor-like signals---their pooled \IC{} significance
reflects stable cross-sectional information in the embedding
structure---while the synthesis compresses these factors into a
concentrated timing signal, the ordinal recommendation, whose date-level
\IC{} reaches formal significance on \spx{} ($p=0.024$) and whose most
confident tier outperforms random selection at $p=0.003$.  No single
agent accounts for either result independently; the edge appears to arise
from their complementary, context-sensitive combination.  The \spone{}
results are directional throughout but underpowered due to the small
per-date sample (${\sim}19$ stocks).

\subsection{Sector Heterogeneity and Weight Drift}
\label{sec:sector}

\begin{figure}
\centering
\includegraphics[width=\textwidth]{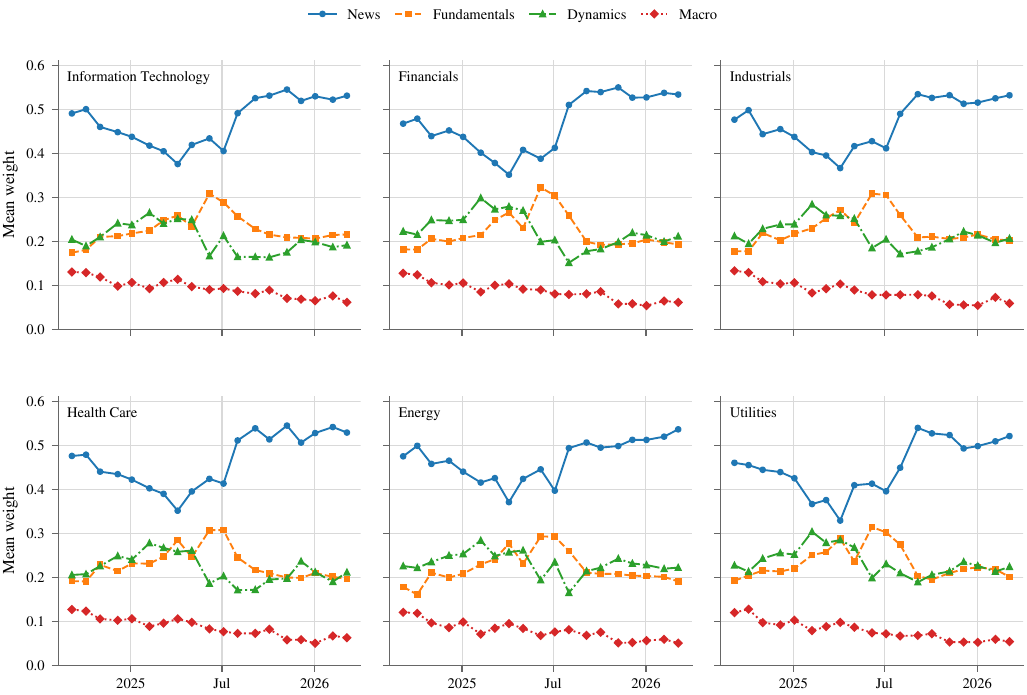}
\caption{%
  Sector mean NNLS agent weights over time (\spx{} cohort).  Each panel shows one sector; lines represent the four agent weights (News, Fundamentals, Dynamics, Macro).  Sectors show persistent differences in which agents dominate consistent with semantic context-adaptivity of the synthesis agent.}
\label{fig:sector_weights}
\end{figure}

Figure~\ref{fig:sector_weights} plots sector mean NNLS weights over time.
There is visible sector-level variation in which agent dominates the thesis:
Information Technology exhibits the highest Macro weight across all sectors
($\bar{w}_{\mathrm{Macro}}=0.093$), while
Energy and Utilities show elevated Dynamics weight---reflecting the
importance of price-momentum signals in commodity-sensitive and
rate-sensitive sectors.  Fundamentals weight is most elevated in Health Care,
with differences across sectors modest but persistent.

The Spearman correlation between reconstruction residual and sector-date centroid drift is $-0.325$ ($p=2.9\times10^{-6}$ pooled) for \spx{}, indicating that dates with higher attribution residuals correspond to sectors undergoing more rapid conceptual shift.  A permutation test on the number of distinct monthly IC-winners by sector returns $p=0.36$, confirming that no single agent consistently dominates across sectors and dates beyond what chance would produce. 
Figure~\ref{fig:best_agent} shows which agent achieves the highest \IC{} on each \spx{} date. The dominant agent rotates across the sample, consistent with the synthesis agent adaptively weighting agents according to prevailing sector conditions and market regime.

\paragraph{Three views of the same rotation:}
The sector composition of strong-buy selections
(Figure~\ref{fig:sector_dist}) rotates in parallel with the best-agent
timeline (Figure~\ref{fig:best_agent}) and the Macro-driven episodes
identified in Table~\ref{tab:macro_episodes}.  The Financials-dominated
Sep 2024 -- Feb 2025 window ($\sim$25\% share vs 15.1\% in the universe)
coincides with Macro-led dates around the Fed easing cycle and the US
election; the Info Tech and Health Care pivot from Mar 2025 onward
corresponds to the post-``Liberation Day'' regime shift and the
transition to Fundamentals- and News-led dates; and the Jun--Aug 2025
collapse of Financials weight alongside the Health Care and Industrials
surge coincides with the recession-concerns rotation captured by
Macro's Aug 2025 leading date.  Sector rotation, agent rotation, and
the macro calendar are three views of the same underlying adaptive
integration: the synthesis agent shifts its selection pool and its
internal agent weighting in tandem as the market regime evolves.

\section{Discussion and Limitations}
\label{sec:discussion}

\paragraph{What the evidence supports:}
The Monte Carlo test provides clear evidence that \msai{}'s strong-buy
selections outperform random same-sized portfolios on the \spx{} cohort,
and the NNLS attribution shows this outperformance cannot be attributed
to any single agent.  The ordinal score's \emph{date-level} \IC{} is
statistically significant on \spx{}, confirming that the recommendation
label carries genuine cross-sectional rank information within the
actionable universe.

\paragraph{Strong-buy as a universe filter:}
A practical consequence of the presented results is that the strong-buy
signal can be read as a \emph{filtering mechanism} rather than a
point-in-time alpha forecast.  Because random equal-weight portfolios
drawn from the strong-buy subset outperform random portfolios drawn from
the broader covered universe at the 99.7th percentile, conditioning any
downstream portfolio-construction process on the strong-buy set---whether
equal-weighting, risk-parity, optimised mean-variance, or a factor
overlay---inherits a universe with better-than-random expected returns by
construction.  The system therefore need not replace existing portfolio
construction methods to add value; it can act upstream of them, shrinking
the investable universe to a candidate pool with improved
ex-ante properties.

\paragraph{Scope of the statistical claims:}
The strongest statistical results in this paper---the \spx{} Monte
Carlo test ($p=0.003$) and the ordinal score's date-level \IC{}
($p=0.024$)---rest on 19 monthly observations and should be interpreted
as evidence from a specific live period rather than asymptotic proof.
Several secondary findings remain directional at the available sample
sizes: the \spone{} Monte Carlo result is consistent with \spx{} but
does not reach formal significance ($p=0.17$, driven by the small
${\sim}10$-stock average selection).  Pooled \IC{}
figures are reported directionally only, as noted in
Section~\ref{sec:ic}, because within-date cross-sectional dependence
makes them unsuitable for conventional significance testing.  In
combination, these scope conditions define what longer time series and
broader universes will be needed to establish; they do not weaken the
primary \spx{} results within the tested window.

\begin{figure}
\centering
\begin{subfigure}[b]{0.46\textwidth}
  \centering
  \includegraphics[width=\textwidth]{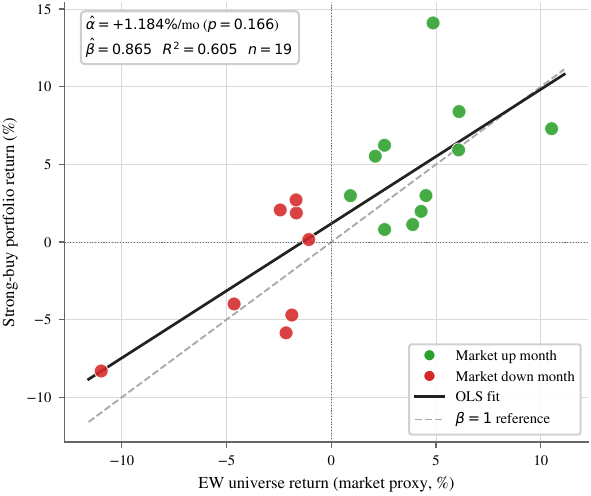}
  \caption{%
    Jensen's $\alpha$ regression: monthly strong-buy portfolio return vs
    EW universe return (market proxy, \spx{} cohort, 19 dates).
    $\hat{\beta}=0.865$ (below unity) and $\hat{\alpha}=+1.18\%$/month
    ($p=0.17$; $R^2=0.60$).  Green/red points indicate up/down market
    months; dashed line is the $\beta=1$ reference.}
  \label{fig:beta_regression}
\end{subfigure}
\hfill
\begin{subfigure}[b]{0.50\textwidth}
  \centering
  \includegraphics[width=\textwidth]{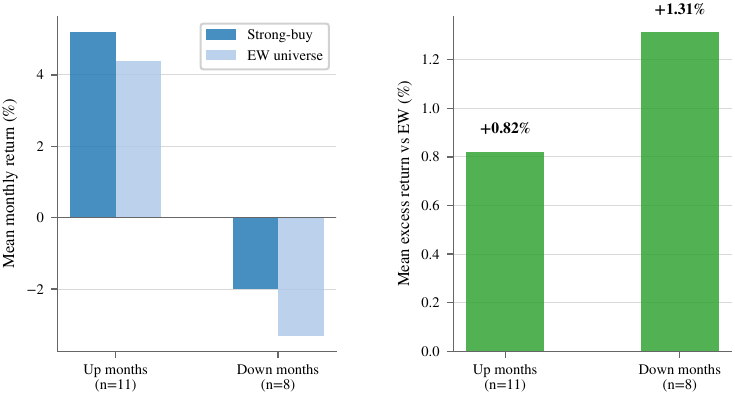}
  \caption{%
    Conditional returns in up-market (11 months) and down-market
    (8 months) periods.  Left: absolute monthly returns for the
    strong-buy portfolio and EW universe.  Right: excess return vs EW.
    The system generates larger excess return in down months
    ($+1.31\%$) than up months ($+0.82\%$), inconsistent with a
    high-beta or momentum explanation.}
  \label{fig:cond_returns}
\end{subfigure}
\caption{%
  Market beta robustness checks (\spx{} cohort).  Neither the below-unity
  portfolio beta ($\hat{\beta}=0.865$) nor the stronger down-market alpha
  preservation is consistent with a pure risk-loading explanation for the
  strong-buy outperformance.}
\label{fig:beta_checks}
\end{figure}

\paragraph{Market beta and downside protection:}
A natural alternative explanation for the strong-buy outperformance is
that the system mechanically selects higher-beta names that benefit from a
broadly risk-on market environment.  We address this directly using the
EW universe return as a within-sample market proxy.
Regressing monthly strong-buy portfolio returns against monthly EW returns
across the 19 \spx{} dates (Figure~\ref{fig:beta_regression}) yields
$\hat{\beta}=0.865$ and a Jensen's $\hat{\alpha}=+1.18\%$/month
($t=1.45$, $p=0.17$; annualised $+14.2\%$, $R^2=0.60$).
The below-unity portfolio beta rules out a pure high-beta explanation; the
alpha is economically large but remains formally underpowered at $T=19$.
Individual strong-buy stock betas (computed in-sample vs the EW proxy)
average $1.06$ against a full-universe mean of $1.00$, confirming that
the selected stocks are only marginally above-average in market sensitivity.

The conditional return breakdown (Figure~\ref{fig:cond_returns}) reinforces
this interpretation.  In the 11 up-market months the strong-buy portfolio
earns $+5.22\%$ against a $+4.39\%$ EW benchmark ($+0.82\%$ excess); in
the 8 down-market months it earns $-2.00\%$ against $-3.32\%$
($+1.31\%$ excess).  The system preserves \emph{more} alpha in adverse
months than in rising ones---the opposite of what a high-beta or momentum
overlay would produce.  Both conditional alphas are positive and
economically meaningful; with only 8 down-month observations formal
significance is not reached ($p=0.28$, one-sample $t$-test).  Taken
together, the below-market beta and the stronger down-market alpha
preservation are inconsistent with a beta-loading explanation and are
consistent with the selection being driven by stock-specific quality
signals rather than systematic risk exposure.

\paragraph{Sell-side pattern:} Sell and strong-sell stocks earn positive average returns over this period. This may reflect the broadly risk-on equity environment: stocks flagged as fundamentally weak may be high-beta names that outperform in a rising market, and as noted in Section~\ref{sec:risk}, short-squeeze dynamics in high-short-interest names can further distort bearish signals over short horizons.  With fewer than 470 sell-side observations total we cannot distinguish these effects from model miscalibration.

It is important to note that evaluating the actionability of sell-side signals requires data that lie outside the scope of this study.  Effective implementation of short positions depends on
\textit{short interest and short ratios} (to assess crowding and squeeze risk), borrow availability (since many fundamentally weak or small-float names are difficult or expensive to borrow), and liquidity metrics (bid-ask spreads, average daily volume) to ensure positions can be entered and exited without material market impact. The absence of these inputs means the sell-side results reported here should be read as a data-scope limitation rather than a verdict on model quality.

\paragraph{Market regime:} The \spx{} period (Sep 2024--Mar 2026) was characterised by generally positive equity returns.  The beta analysis above (Figure~\ref{fig:beta_checks}) provides partial reassurance: the system generates larger excess returns in down-market months than up-market months, and its portfolio beta is below unity. However, whether the outperformance persists in a sustained bear market is unknown. The \spone{} period covers an additional 16 months (May 2023--Aug 2024) that includes more varied conditions, and the directional result holds, but this does not constitute full out-of-sample validation across regimes.

\section{Conclusion}
\label{sec:conclusion}

We show that the strong-buy signals produced by \msai{}, a deployed
multi-agent LLM equity system, outperform both a passive equal-weight
benchmark of the covered universe (approximating RSP on \spx{} and EQWL
on \spone{}) and random same-sized portfolios drawn from the same universe
on the same dates.  On the \spx{} cohort the strong-buy portfolio delivers
a $+25$ percentage-point compound-return advantage over the EW benchmark
and ranks at the 99.7th percentile of 10{,}000 Monte Carlo null portfolios
($p=0.003$); on the \spone{} cohort the excess over EQWL is
$+30$ percentage points compounded, with the directional pattern holding
but formal significance not reached given the smaller average position
count per date.

NNLS attribution of thesis embeddings onto four agent embeddings reveals
that agent contributions are heterogeneous and context-dependent: no single
agent dominates across all dates and sectors.  Fundamentals leads on \spx{},
Macro on \spone{}, while Dynamics leads on 5 of 19 \spx{} dates despite a
negative aggregate pooled \IC{}, consistent with momentum being a
regime-conditional rather than persistent signal.  The ordinal score's
date-level \IC{} is statistically significant on \spx{}, confirming that
the synthesis agent's label carries genuine cross-sectional rank
information within the actionable buy+strong-buy universe.  Taken together,
the portfolio outperformance and the \IC{} results are consistent with the
synthesis agent acting as an adaptive integrator that exploits each
agent's comparative advantage by sector and market regime---an attribution
structure that would be invisible if one evaluated only the discrete output
label.

These results suggest that multi-agent LLM equity systems can identify new sources of alpha that are not captured by traditional quantitative factor models, and that interpreting such systems through their internal reasoning structure may reveal economically meaningful signal that would otherwise be invisible.  Future work should extend the time series, test the approach on larger and more diverse equity universes (including international and mid-cap coverage), evaluate robustness across market regimes, and examine whether the attribution patterns shift during periods when the system's edge is absent.

\bibliographystyle{plainnat}
\bibliography{references}

\appendix
\newpage
\section{Monte Carlo Results by Date}
\label{app:mc_dates}

Table~\ref{tab:mc_sp500} reports the month-by-month
Monte Carlo results for the \spx{} cohort.
Each row shows the number of strong-buy picks, the actual equal-weight
return, the null mean, the excess return, and the percentile rank of the
actual return in the null distribution.

\begin{table}
\centering
\caption{Monte Carlo results by date --- \spx{} cohort.
  Pct: percentile rank of actual return in the null distribution of
  10{,}000 random same-sized portfolios.}
\label{tab:mc_sp500}
\begin{tabular}{lrcccc}
\toprule
Date & $n_{\mathrm{SB}}$ & Actual (\%) & Null mean (\%) & Excess (\%) & Pct \\
\midrule
2024-09 & 48 & $+5.93$ & $+6.07$ & $-0.14$ & 48 \\
2024-10 & 57 & $+0.16$ & $-1.09$ & $+1.25$ & 91 \\
2024-11 & 55 & $+8.40$ & $+6.11$ & $+2.28$ & 91 \\
2024-12 & 74 & $-3.99$ & $-4.64$ & $+0.65$ & 88 \\
2025-01 & 41 & $+6.23$ & $+2.55$ & $+3.68$ & $>99$ \\
2025-02 & 55 & $-5.85$ & $-2.16$ & $-3.69$ & $<1$ \\
2025-03 & 30 & $-8.30$ & $-11.00$ & $+2.70$ & 97 \\
2025-04 & 18 & $+7.30$ & $+10.55$ & $-3.25$ &  9 \\
2025-05 & 25 & $+2.99$ & $+4.55$ & $-1.55$ & 18 \\
2025-06 & 19 & $+1.98$ & $+4.29$ & $-2.31$ &  7 \\
2025-07 & 21 & $+1.87$ & $-1.66$ & $+3.53$ & 98 \\
2025-08 & 28 & $+1.13$ & $+3.89$ & $-2.76$ &  4 \\
2025-09 & 34 & $+5.53$ & $+2.07$ & $+3.45$ & 98 \\
2025-10 & 26 & $+2.71$ & $-1.68$ & $+4.40$ & $>99$ \\
2025-11 & 32 & $+0.81$ & $+2.55$ & $-1.75$ &  6 \\
2025-12 & 25 & $+2.99$ & $+0.92$ & $+2.07$ & 97 \\
2026-01 & 20 & $+14.11$ & $+4.89$ & $+9.22$ & $>99$ \\
2026-02 & 36 & $-4.70$ & $-1.87$ & $-2.83$ &  3 \\
2026-03 & 22 & $+2.07$ & $-2.45$ & $+4.51$ & $>99$ \\
\midrule
\textit{Mean} & 35.1 & $+2.18$ & $+1.15$ & $+1.02$ & 60.7 \\
\bottomrule
\end{tabular}
\end{table}

\end{document}